\documentclass[review]{elsarticle}

\usepackage{lineno,hyperref}
\modulolinenumbers[5]

\journal{Journal of \LaTeX\ Templates}

\begin{document}

\begin{frontmatter}

\title{
\vskip-1.5cm{\baselineskip14pt\rm
\centerline{\normalsize DESY 16-020\hfill ISSN 0418-9833}
\centerline{\normalsize January 2016\hfill}}
\vskip1.5cm
Counting master integrals: Integration by parts vs.\ functional
equations}

\author{Bernd A. Kniehl\corref{mycorrespondingauthor}}
\cortext[mycorrespondingauthor]{Corresponding author}
\ead{kniehl@desy.de}
\author{Oleg V. Tarasov\fnref{myfootnote}}
\fntext[myfootnote]{On leave of absence from Joint Institute for Nuclear
Research, 141980 Dubna (Moscow Region), Russia.}
\ead{oleg.tarasov@desy.de}
\address{II. Institute f\"ur Theoretische Physik, Universit\"at Hamburg,
Luruper Chaussee 149, 22761 Hamburg, Germany}

\begin{abstract}
We illustrate the usefulness of functional equations in establishing
relationships between master integrals under the integration-by-parts reduction
procedure by considering a certain two-loop propagator-type diagram as an
example.
\end{abstract}

\begin{keyword}
Two-loop sunset diagram \sep 
Recurrence relations \sep
Functional equations \sep
Multiloop calculations

\PACS 02.30.Gp \sep 02.30.Lt \sep 11.15.Bt \sep 12.38.Bx
\end{keyword}

\end{frontmatter}


An adequate theoretical interpretation of the increasingly precise data
collected by the experiments at the CERN Large Hadron Collider and elsewhere
necessitates advanced technologies for the calculation of radiative
corrections, which typically depend on several different mass scales. 
Feynman diagrams involving quantum loops may be reduced to so-called
master integrals via dedicated algorithms, such as integration by parts (IBP)
\cite{Tkachov:1981wb,Chetyrkin:1981qh}.
The evaluation of the master integrals often turns out to be a bottleneck of
the entire theoretical analysis, the more if many different mass scales are
involved.
Any method to reduce the number of master integrals of a given set of
Feynman diagrams is, therefore, highly welcome.
Recently, relationships between master integrals of the two-loop sunset
diagram were found in Refs.~\cite{Kalmykov:2011yy,Kniehl:2012hn}.
In the present paper, a new relationship of this type will be presented,
which is found using functional equations \cite{Tarasov:2008hw}.




The derivation of functional equations for integrals with two and more loops
is much more complicated than in the one-loop case.
In this following, we consider two-loop propagator-type integrals.
Two-loop integrals differ by the number of internal lines.
According to the algorithm of Refs.~\cite{Tarasov:2008hw,Tarasov:2011zz},
functional equations may be obtained from recurrence relations connecting
two-loop integrals with different numbers of lines.
The most complicated integrals in such a functional equation may be eliminated
by an appropriate choice of four-momenta and masses.

\begin{figure}[h]
\begin{center}
\includegraphics[scale=1.1]{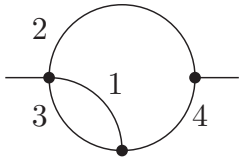}
\end{center}
\caption{\label{fig:one}%
Two-loop diagram corresponding to integral $V^{(d)}_{1111}$.}
\end{figure}
Let us consider the following two-loop propagator-type integral with four
internal lines: 
\begin{eqnarray}
\lefteqn{V^{(d)}_{\nu_1\nu_2\nu_3\nu_4}(m_1^2,m_2^2,m_3^2,m_4^2;q^2)
=\frac{1}{\left(i \pi^{d/2}\right)^2}}
\nonumber\\
&&\hspace{-1.2cm}{}\times
\int\int \frac{d^d k_1 d^dk_2}{[(k_1-k_2)^2-m_1^2]^{\nu_1}
[k_2^2-m_2^2]^{\nu_2}
[(k_1-q)^2-m_3^2]^{\nu_3}[(k_2-q)^2-m_4^2]^{\nu_4}},
\end{eqnarray}
where $d$ is the space-time dimension.
The Feynman diagram corresponding to this integral is shown in
Fig.~\ref{fig:one}.
With the aid of generalized recurrence relations given in
Ref.~\cite{Tarasov:1997kx}, the integral $V^{(d)}_{\nu_1\nu_2\nu_3\nu_4}$ with
arbitrary integers $\nu_j$ may be reduced to the integral $V^{(d)}_{1111}$,
four two-loop integrals with three lines of the type
\begin{eqnarray}
\lefteqn{J^{(d)}_{\nu_1 \nu_2 \nu_3}(m_1^2,m_2^2,m_3^2;q^2)}
\nonumber\\
&=&
\frac{1}{\left(i \pi^{d/2}\right)^2}
\int\int \frac{d^d k_1 d^dk_2}
{[k_1^2-m_1^2]^{\nu_1}[(k_1-k_2)^2-m_2^2]^{\nu_2}
[(k_2-q)^2-m_3^2]^{\nu_3}},
\end{eqnarray}
the product of a one-loop propagator-type integral,
\begin{equation}
I_2^{(d)}(m_1^2,m_2^2;q^2)=
\frac{1}{i\pi^{d/2}}\int\frac{d^d k_1}{(k_1^2-m_1^2)[(k_1-q)^2-m_2^2]},
\end{equation}
with masses $m_2^2$ and $m_3^2$, times a one-loop vacuum-type integral,
\begin{equation}
T_a(m_1^2)=\frac{1}{i\pi^{d/2}}\int\frac{d^d k_1}{k_1^2-m_1^2},
\end{equation}
and products of one-loop vacuum-type integrals with different masses.

In Ref.~\cite{Tarasov:1997kx}, several generalized recurrence relations 
for the integral $V^{(d)}_{\nu_1\nu_2\nu_3\nu_4}$ were presented. 
One of these relations, namely the one in Eq.~(55) therein, reads:
\begin{eqnarray}
m_4^2V^{(d)}_{1112}&=& 
  \frac{(d-3)  u_{624}u_{246} \Delta_{134}
 + (d-3) u_{314}u_{134}\Delta_{246} +   (d-4) \Delta_{134} \Delta_{246}}
{2\Delta_{134} \Delta_{246}}
\nonumber
\\
&&{}\times V^{(d)}_{1111}
+\frac{u_{624}  \Delta_{134}-u_{314} \Delta_{246}}
{\Delta_{134}\Delta_{246}}m_3^2
J^{(d)}_{112}(m_1^2,m_2^2,m_3^2;q^2)
\nonumber
\\
&&{}+ \frac{u_{624} \Delta_{134}-u_{134}\Delta_{246}}
{  \Delta_{134} \Delta_{246}}
m_1^2J^{(d)}_{211}(m_1^2,m_2^2,m_3^2;q^2)
\nonumber
\\
&&{}+\frac{2m_2^2(q^2-m_2^2)}{ \Delta_{246}}
J^{(d)}_{121}(m_1^2,m_2^2,m_3^2,q^2)
\nonumber
\\
&&{}- \frac{u_{624} (3 d-8)}{2  \Delta_{246}}
J^{(d)}_{111}(m_1^2,m_2^2,m_3^2;q^2)
\nonumber
\\
&&{}+ \frac{(d-2)u_{624}}{2  \Delta_{246}}
 J_{111}(m_1^2,m_3^2,m_4^2;0)
\nonumber
\\
&&{}+\frac{(d-2)}{2\Delta_{134}}\left[u_{314}T^{(d)}_1(m_3^2)
+u_{134}T^{(d)}_1(m_1^2)\right]I_2^{(d)}(m_2^2,m_4^2;q^2).
\label{eq:v1112}
\end{eqnarray}
where $u_{ijk}=m_i^2-m_j^2-m_k^2$ and
$\Delta_{ijk}=-u_{ijk}(u_{jik}+u_{kij})-u_{jik}u_{kij}$.
According to the algorithm of Ref.~\cite{Tarasov:2008hw} to obtain functional
equation, one has to eliminate from Eq.~(\ref{eq:v1112}) the integrals
$V_{1111}^{(d)}$ and $V_{1112}^{(d)}$ by an appropriate choice of four-momentum and
masses.
For $m_4=0$, the left-hand side of Eq.~(\ref{eq:v1112}) vanishes, so that we
obtain an expression for the integral $V_{1111}^{(d)}$ in terms of integrals
with lesser numbers of lines, namely,
\begin{eqnarray}
V_{1111}^{(d)}&=&
\frac{2m_1^2(q^2+u_{312} )}{(d-2)(m_3^2-m_1^2)(q^2-m_2^2)}
J_{211}(m_1^2,m_2^2,m_3^2;q^2)
\nonumber
\\
&&{}-\frac{2m_3^2 (q^2+u_{123})}{(d-2)(m_3^2-m_1^2)(q^2-m_2^2)}
J_{112}(m_1^2,m_2^2,m_3^2;q^2)
\nonumber
\\
&&{}+\frac{4m_2^2}{(d-2)(q^2-m_2^2)}
J_{121}(m_1^2,m_2^2,m_3^2;q^2)
\nonumber
\\
&&{}-\frac{(3d-8)}{(d-2)(q^2-m_2^2)}
J_{111}(m_1^2,m_2^2,m_3^2;q^2)
\nonumber
\\
&&{}+\frac{1}{q^2-m_2^2}
J_{111}(m_1^2,m_3^2,0;0)
\nonumber
\\
&&{}+\frac{1}{m_1^2-m_3^2}
\left[T^{(d)}_1(m_1^2)-T^{(d)}_1(m_3^2)\right]I_2^{(d)}(m_2^2,0;q^2).
\label{eq:v1111}
\end{eqnarray}
After multiplying Eq.~(\ref{eq:v1111}) with the factor $q^2-m_2^2$ and then
setting $q^2=m_2^2$, the contribution proportional to the integral
$V^{(d)}_{1111}$ drops out, and obtain the following relationship: 
\begin{eqnarray}
0&=&2m_1^2J^{(d)}_{211}(m_1^2,m_2^2,m_3^2;m_2^2)
  +4m_2^2J^{(d)}_{121}(m_1^2,m_2^2,m_3^2;m_2^2)
\nonumber
\\
&&{}+2m_3^2J^{(d)}_{112}(m_1^2,m_2^2,m_3^2;m_2^2)
-(3d-8)J^{(d)}_{111}(m_1^2,m_2^2,m_3^2;m_2^2)
\nonumber
\\
&&{}+(d-2)J_{111}^{(d)}(m_1^2,m_3^2,0;0).
\label{qq=mm2}
\end{eqnarray}
Equation~(\ref{qq=mm2}) connects two-loop propagator-type integrals with
different kinematics.
The analytic expression for the integral $J_{111}^{(d)}(m_1^2,m_2^2,m_3^2;0)$
in terms of the Gauss hypergeometric function ${}_2F_1$ presented in
Ref.~\cite{Davydychev:1992mt} is considerably simpler than the analytic
expressions for the integrals $J_{111}^{(d)}$ and $J_{211}^{(d)}$ with external
momentum square being different from zero.
It is interesting to notice that the Cayley--Menger determinant
\begin{eqnarray}
D_{123}&=&[q^2-(m_1+m_2+m_3)^2][q^2-(m_1-m_2+m_3)^2]
\nonumber \\
&&{}\times [q^2-(m_1+m_2-m_3)^2][q^2-(m_1-m_2-m_3)^2]
\end{eqnarray}	
for this kinematics is different from zero:
\begin{equation}
\left.D_{123}\right|_{q^2=m_2^2}=
(m_1^2-m_3^2)^2 [(m_1^2-m_3^2)^2 +8m_2^2(2m_2^2-m_1^2-m_3^2)].
\end{equation}
Thus, Eq.~(\ref{qq=mm2}) is a clear illustration that the number of nontrivial
basis integrals, as predicted by IBP, may re reduced not only if $D_{123}=0$ or
one mass is zero as was observed in Ref.~\cite{Tarasov:1997kx}, but also for
other values of four-momentum momentum square and masses.
One possible interpretation is that the total number of basis integrals arising
from the IBP reduction of the integral
$J^{(d)}_{\nu_1 \nu_2 \nu_3}(m_1^2,m_2^2,m_3^2;m_2^2)$
with arbitrary integer powers of propagators remains the same, but that one
nontrivial integral may be replaced by simpler one.

For the particular case when $m_1=0$ and $m_2^2=m_3^2=m^2$, the reduction of
the number of basis integrals was observed in Ref.~\cite{Davydychev:2000na}.
For this kinematics, Eq.~(\ref{qq=mm2}) yields
\begin{equation}
J^{(d)}_{211}(m^2,0,m^2;m^2)
=\frac{3d-8}{6m^2} J^{(d)}_{111}(m^2,0,m^2;m^2)
-\frac{d-2}{6m^2} J_{111}^{(d)}(m^2,0,0;0),
\label{mm1=0}
\end{equation}
so that, instead of two nontrivial integrals, only one nontrivial basis
integral,\break $ J^{(d)}_{111}(m^2,0,m^2;m^2)$, remains.

Putting $m_3=0$ in Eq.~(\ref{qq=mm2}), we recover Eq.~(9) in
Ref.~\cite{Kalmykov:2011yy}, which was obtained there as a special case via
differential reduction
\cite{Kalmykov:2006pu,Kalmykov:2009tw,Bytev:2009mn,Bytev:2009kb,%
Kalmykov:2010gb,Kalmykov:2010xv,Yost:2011hk,Bytev:2011ks,Yost:2011wk,%
Kalmykov:2012rr,Bytev:2012ud,Bytev:2013gva,Bytev:2013bqa}.
Another generalization of Eq.~(9) in Ref.~\cite{Kalmykov:2011yy} was obtained
in Ref.~\cite{Kniehl:2012hn} using IBP in connection with an effective
propagator mass \cite{Kniehl:2005yc}.

We would like mention that Eq.~(\ref{qq=mm2}) connects integrals of different
mass assignments.
Such integrals may arise from rather different Feynman diagrams.
Relationships of this type may be very useful, e.g., for proving the gauge
independence of radiative corrections to physical observables.

In conclusion, functional equations \cite{Tarasov:2008hw,Tarasov:2011zz}
provide a powerful tool for disclosing hidden relationships between what appear
to be master integrals upon standard applications of the IBP reduction
procedure \cite{Tkachov:1981wb,Chetyrkin:1981qh}.
Similar relationships have previously been revealed using differential
reduction \cite{Kalmykov:2011yy} and a nonstandard variant of the IBP
reduction procedure  implemented with propagator masses to
be integrated over \cite{Kniehl:2012hn}.

\section*{Acknowledgments}

This work was supported by the German Research Foundation DFG through the
Collaborative Research Center SFB~676 {\it Particles, Strings and the Early
Universe: the Structure of Matter and Space-Time}.


\bibliography{hep}

\end{document}